# Diffraction of magnetic field in a device of circular rings

Y. Contoyiannis[(1)] and M. Kampitakis[(2)]

(1) Department of Electric-Electronics Engineering, West Attica University, 250 Thivon and P. Ralli, Aigaleo, Athens GR-12244, Greece (email: yiaconto@uniwa.gr)
(2) Hellenic Electricity Distribution Network Operator SA, Network Major Installations Department, 72 Athinon Ave., N.Faliro GR-18547, Greece (email: m.kampitakis@deddie.gr)

**Abstract** : A device consisted of a set of circular rings, the centers of which lie on an axis, behaves like a solenoid when the ratio of its radius and distance between two successive rings is greater than one. As this ratio decreases, the device deviates a lot from the solenoid behavior. In the same way, a diffraction phenomenon for magnetic field appears when currents of random direction flow through the rings. This phenomenon demonstrates a critical behavior. Thus an extension of diffraction phenomenon can be done beyond the classical wave diffraction. A possible application of this device could be the diffraction of electronic beams.

In one of our previous works [1] we had presented a device consisted of a set of similar rings with radius α, where its centers lie on an vertical axis to their plane and the distance between two successive centers is d (Fig. 1). If we supply this device with currents of random direction +I, -I then when we decrease the number of rings per unit length, we observe a transition from a continuous to a stratified spectrum of magnetic field (magnetic induction) values at the centers of the rings on the symmetry axis.

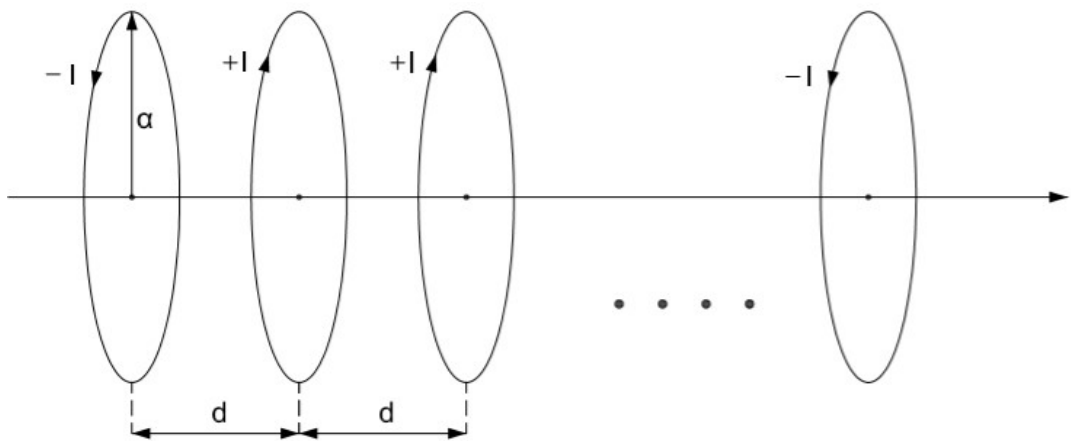

**Fig.1** *The proposed device of circular rings*

The aim of our present work is to investigate the mechanism of this transition. Thus we could answer the question if the stratified spectrum of magnetic field in the case where the currents of rings have a random direction, refer to a form of diffraction phenomenon connected with the magnetic field behavior.

The diffraction of waves is a phenomenon where an obstacle or a narrow slot is in the path of the wave propagation. Although diffraction is a phenomenon of waves, great similarities and proportions appear in the magnetic field produced by this device when its currents have random directions. It is interesting if in this way, diffraction could be extended in other physical phenomena.

Until recently, diffraction phenomena can appear in a high electron beam, which is accelerated in a microwave electron gun. After passing through the solenoid lens, the electron beam diffracts on the sample and finally produces the diffraction pattern on the fluorescent screen (Ultrafast Electron Diffraction) [2,3]. In our case, an interesting application is that our device achieves very high spatial resolution due to the strong diffracted magnetic field.

We consider a set from similar circular rings, the centers of which lie on an axis of symmetry. The currents which flow through these rings create magnetic fields. The magnetic field on the axis due to each ring is given by the Biot-Savart law as [4,5]:

$$B_x = \frac{\mu_o I a^2}{2(x^2 + a^2)^{3/2}} \quad (1)$$

where x is the distance from the center of the ring, I is the current and α the radius of the ring. In the discrete version of our model we have that $x = nd$ where n = 0, 1, 2, …. while the case n = 0 corresponds to the distance between the ring and itself. Therefore, the magnetic field in the center of each ring is due to its magnetic field and the contribution from all other rings.

Letting x = nd, equation 1 can be written as :

$$B_n = \frac{\mu_o I a^2}{2(d^2 n^2 + a^2)^{3/2}} \quad (2)$$

Assuming that the number of rings is N and the universal constant $\mu_o = 1$, then the magnetic field of the k-ring (k=1,2, …, N) is of the following form :

$$B_k = \frac{a^2}{2}\left\{\sum_{n=0}^{k-1} \frac{I_{k-n}}{[(nd)^2 + a^2]^{3/2}} + \sum_{n=0}^{N-k-1} \frac{I_{n+1+k}}{[(nd+d)^2 + a^2]^{3/2}}\right\} \quad (3)$$

The linear density of rings is $\rho = \frac{N}{L}$ where $L = (N-1)d$ is the length of the system of rings. Usually the number of rings is large enough so that $N \gg 1$ and therefore we can write $\rho \approx \frac{N}{Nd} = \frac{1}{d}$.

Letting $c = \frac{a}{d}$ in (3) we have:

$$B_k = \frac{1}{2} I a^2 d^{-3}\left\{\sum_{n=0}^{k-1} \frac{1}{(n^2 + c^2)^{3/2}} + \sum_{n=0}^{N-k-1} \frac{1}{[(n+1)^2 + c^2]^{3/2}}\right\} \quad (4)$$

An analytical estimation for the field $B_k$ from eq. 4 is possible if $= N \to \infty$. Then $B_N$ is the magnetic field (magnetic induction) in the edge of the device which is equal to $B_\mu/2$ [5], where $B_\mu$ is the magnetic field on the axis in the interior of device. So we can write :

$$B_\mu = I\alpha^2 d^{-3} \sum_{n=0}^{\infty} \frac{1}{(n^2 + c^2)^{3/2}} = I\alpha^2 d^{-3} \left[ \frac{1}{c^2} + \sum_{n=1}^{\infty} \frac{1}{(n^2 + c^2)^{3/2}} \right] \quad (5)$$

It is easily proved that the positive sum above converges, namely

$$\sum_{n=1}^{\infty} \frac{1}{(n^2 + c^2)^{3/2}} = S$$

S can be estimated according to Maclaurin - Cauchy test :

$$\int_N^\infty f(x)dx \leq \sum_{n=N}^{\infty} f(n) \leq f(N) + \int_{n=N}^{\infty} f(x)dx$$

for $N = 1$. We consider the function $f(x) = \frac{1}{(x^2+c^2)^{3/2}}$ and estimate the integral $A = \lim_{r \to \infty} \int_1^r f(x)dx$. From the table of integrals [6] we have that

$$\int \frac{1}{(x^2 + c^2)^{3/2}} dx = \frac{x}{c^2\sqrt{x^2 + c^2}}$$

As a result:

$$A = \frac{1}{c^2} - \frac{1}{c^2\sqrt{1 + c^2}} \quad (6)$$

The quantities A and S do not exactly equal each to other but an uncertainty exists according to **Maclaurin-Cauchy test**. Thus the sum has a value inside the interval: A < S < A+δ, where $\delta = f(1)$. Therefore, S can be written as :

$$\frac{1}{c^2} - \frac{1}{c^2\sqrt{1 + c^2}} < S < \frac{1}{c^2} - \frac{1}{c^2\sqrt{1 + c^2}} + \frac{1}{(1 + c^2)^{3/2}} \quad (7)$$

Finally the magnetic field $B_\mu$ is expressed as :

$$B_\mu = I\alpha^2 d^{-3} \left( \frac{1}{c^3} + S \right) \quad (8)$$

We consider the following cases:

(a) $c \gg 1$

In this case the quantity $\delta = \frac{1}{(1+c^2)^{3/2}} \to 0$ and

$$S = \frac{1}{c^2}\left(1 - \frac{1}{\sqrt{1 + c^2}}\right) \quad (9)$$

Substituting eq. (9) in eq. (8) we obtain :

$$B_\mu = I\alpha^2 d^{-3}\left[\frac{1}{c^3} + \frac{1}{c^2}\left(1 - \frac{1}{\sqrt{1+c^2}}\right)\right] = I\alpha^2 d^{-3}\frac{1}{c^2}\left(\frac{1}{c} + 1 - \frac{1}{\sqrt{1+c^2}}\right)$$

$$B_\mu = I\alpha^2 d^{-3}\frac{d^2}{\alpha^2}\left(\frac{1}{c} + 1 - \frac{1}{\sqrt{1+c^2}}\right) = Id^{-1}\left(\frac{1}{c} + 1 - \frac{1}{\sqrt{1+c^2}}\right)$$

Under the condition $c \gg 1$ we have

$$\frac{1}{\sqrt{1+c^2}} \approx \frac{1}{c}$$

and the expression for $B_\mu$ becomes:

$$B_\mu = Id^{-1}\left(\frac{1}{c} + 1 - \frac{1}{c}\right) = Id^{-1} = I\lambda \quad (10)$$

This is the Ampere law which is valid in the solenoid case. We conclude from eq.10 that in the case $\alpha \gg d$, the device of rings behaves as a solenoid. In a plot of $B_\mu$ vs $d$ we shall take a line with slope -1.

(b) $c \ll 1$

In this case the quantity

$$\delta = \frac{1}{(1+c^2)^{3/2}} \to 1$$

and quantity

$$\frac{1}{c^2}\left(1 - \frac{1}{\sqrt{1+c^2}}\right) \to 0$$

Therefore S can be written as (eq. 7):

$$0 < S < 1$$

while $B_\mu$ from eq.8 becomes:

$$B_\mu = I\alpha^2 d^{-3}\frac{1}{c^3} + I\alpha^2 d^{-3}\,S = I\alpha^{-1} + Ic^2 d^{-1}\,S \quad (11)$$

It is obvious from eq.11 that the law of Ampere is not valid anymore and the device does not behave as a solenoid.

(c) c = 1

From eq.8 we deduce that:

$$B_\mu = Id^{-1}(1 + S) \quad (12)$$

while, as a result of eq.7, the possible S-values lie in the interval

$$1 - \frac{1}{\sqrt{2}} < S < 1 - \frac{1}{\sqrt{2}} + \frac{1}{2^{3/2}} \quad (13)$$

We perform a numerical experiment based on an algorithm resulting from eq.(3) letting $a = 5, I = 1$, where we estimate the magnetic field $B_\mu$ vs d as we change the linear density of rings $\lambda = d^{-1}$. A number of rings N = 5000 ensures the convergence of magnetic field (magnetic induction). In Fig.2 we present the results of this numerical experiment.

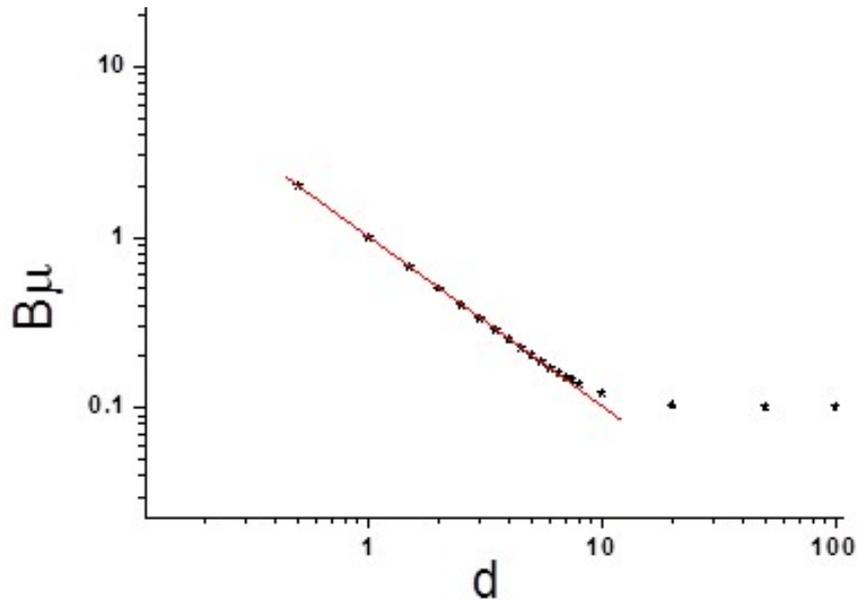

**Fig. 2** *The magnetic field in the interior of the rings set, on the symmetry axis vs the distance between two successive rings.*

As demonstrated in fig 2, the ampere law (10) for $c \gg 1$ ($d \ll a$). We see a continuous transition from linearity to non linearity as c approaches the value $c = 1$ ($d = a$). For $c \ll 1$ a convergence to a constant $B_\mu$-value appears regardless the value of d, verifying in this way eq.11 provided that S takes its minimum value, namely S=0, and therefore the magnetic field becomes :

$$B_\mu = I\alpha^{-1} \quad (14)$$

In Fig. 3 a detail of Fig.2 is shown, as the points desert linearity.

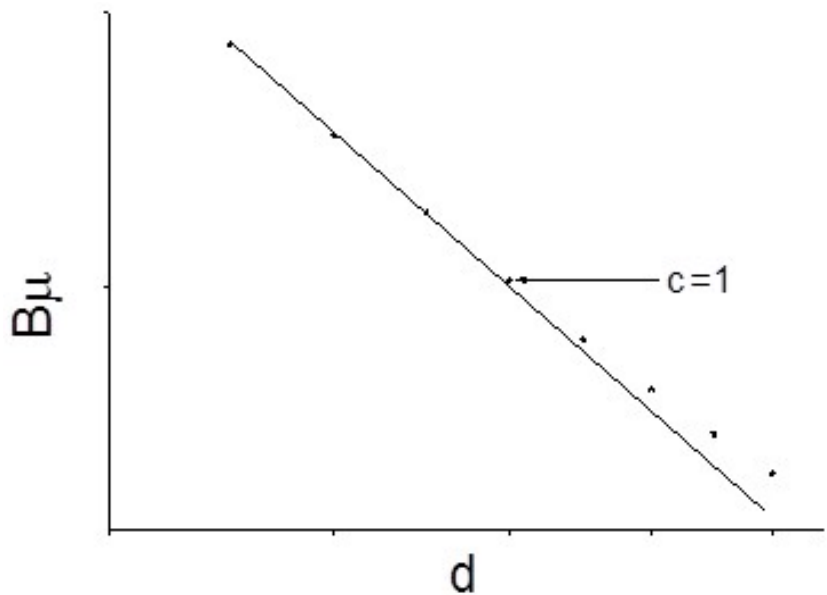

**Fig.3** *A detail from Fig.2 shows that deviation from the linear behavior begins at c = 1.*

As we mentioned above, in [1] we have shown that the ring model gives images that resemble diffraction images when flowing through the rings have currents the same value but random directions. The picture of layering in the ring system appears when the linear density of the rings decreases, keeping the radius constant. The analysis performed in this work helps us to better understand the phenomenon of the fringes we had presented in [1]. In the following figures (4,5) the phenomenon of diffraction for the magnetic field in the interior of the device is demonstrated.

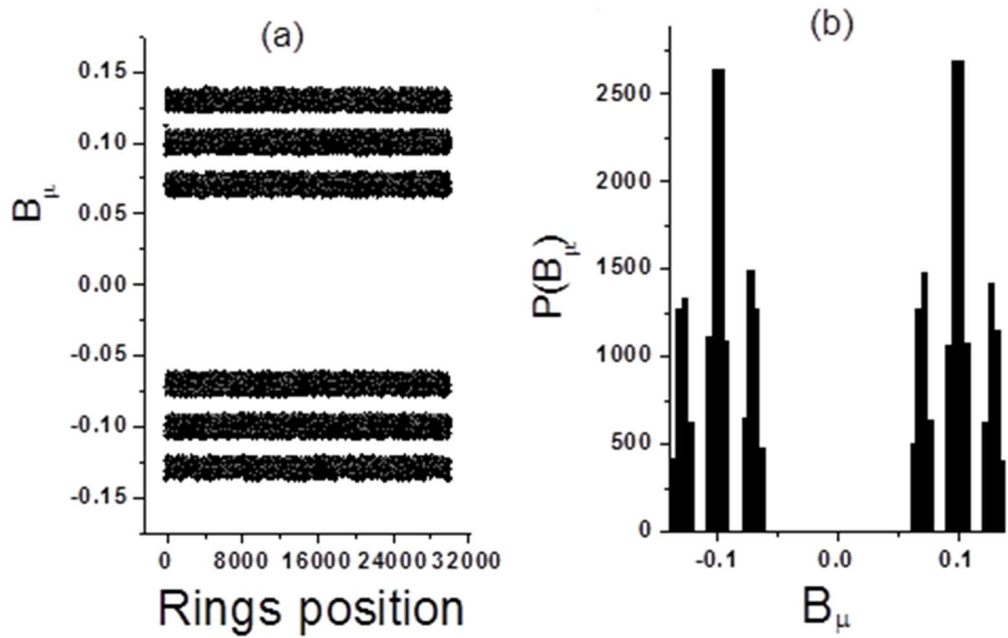

**Fig. 4** *(a) The magnetic field $B_\mu$ vs the rings position for $c = \frac{a}{d} = \frac{5}{8}$ is shown. In the central region a cancelation zone appears for certain values of the magnetic field, while the same phenomenon is present in the other four smaller cancelation zones. (b) The distribution of magnetic field resembles the intensity of the waves diffraction.*

As we can see from Fig.2, we have a continuous transition from a state ($c \gg 1$) where the magnetic field is proportional to the linear density of the rings (Ampere law eq. 10) and the state ($c \ll 1$) where the magnetic field is independent of the linear density (eq.14). From Fig. 3 this transition appears when the value of $c$ is very close to 1. If the diffraction phenomenon as described in Fig. 4 exists, this continuous transition would have to appear in the patterns and should begin to appear when c=1. In Fig. 5 an evolution of the magnetic field and its distribution is presented, as the linear density of the rings decreases up to the critical point ($c \approx 1$) where the pattern of diffraction appears.

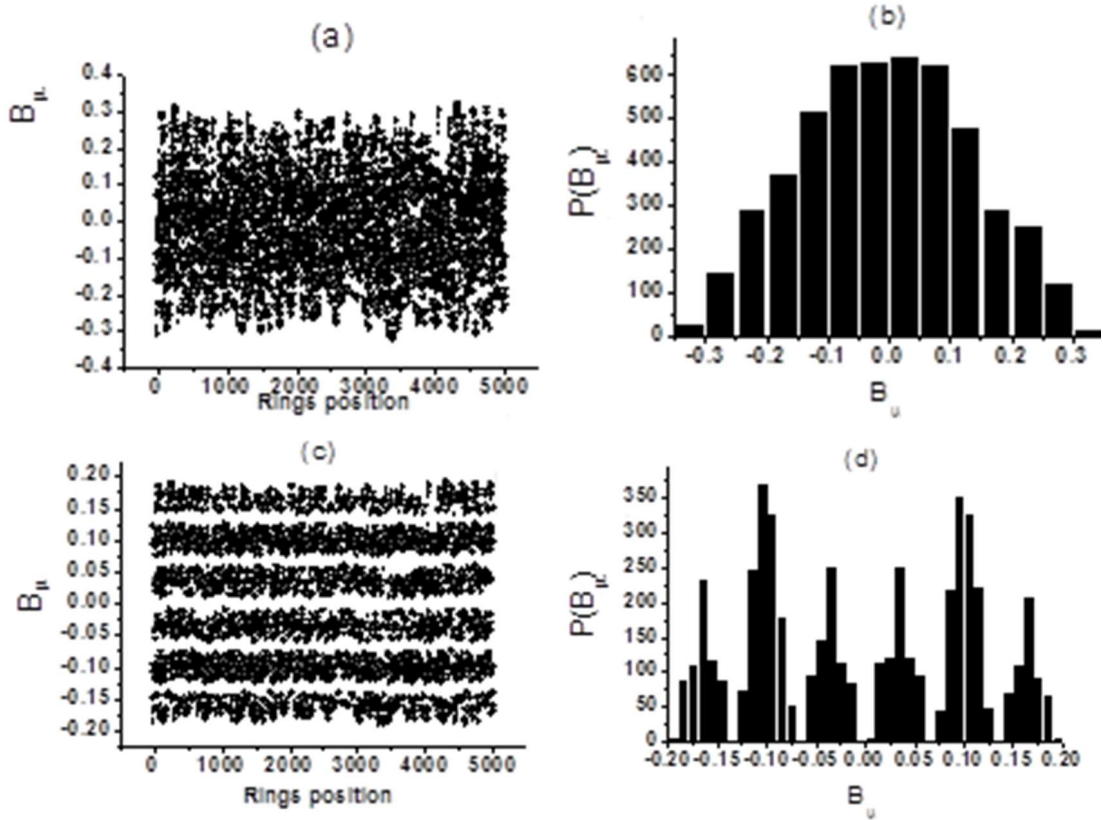

**Fig.5 (a), (b)** *The pattern and the intensity of distribution of the magnetic field in the interior of the rings device when $c = \frac{5}{3} > 1$. Due to the randomness of the currents directions, areas where the magnetic fields cancel (white areas) intersect chaotically with black areas where the magnetic fields have survived.* **(c), (d)** *The appearance of the central region of the cancelation of magnetic field consummates the creation of the diffraction pattern. This happens when $= \frac{5}{5.25} = 0.95$, a value very close to the critical point $c = 1$, where the law of Ampere ceases to apply anymore. As $c$ continues to decrease, the diffraction pattern becomes distinct, as in Fig. 4.*

In the following, we will try to give a qualitative interpretation of the magnetic field diffraction phenomenon based on similarities and differences with corresponding wave quantities.

In the region c≫1 up to c = 1, the high linear density of the rings namely the small distances between the neighboring rings and the greater contribution to the creation of the magnetic fields make the magnetic field dependent on the positions of the rings (fig.5a) . In the region from c = 1 up to c << 1 we have a continuous increase in the distance d of the rings, which means that we are moving towards the independence of the magnetic field from the position of the rings.  When c << 1, the magnetic field is completely independent of d (eq.14). This process of the independence of the magnetic field from the positions of the rings is expressed by stratifying the magnetic field values in a diagram (Figure 4a). One could say that this evolution as we reduce the linear density, is expected. However, the confirmation that the diffraction image is completed at c = 0.95≈1 (Fig. 5c,d) where the Ampere law ceases to apply shows that the evolution of the magnetic field pattern follows the theory we have outlined above. Therefore, the diffraction phenomenon appears as a result of the magnetic field behavior, as the ratio c varies, as it is presented above.

Such diagrams in Fig. 4 have much in common with known diffraction patterns of waves from holes, slits and / or obstacles (here in obstacles) with the ratio $α / λ << 1$, where α is the radius of the hole and λ is the wavelength. We shall then refer to such common features

- In wave diffraction, the images of the wave intensity distribution refer to a horizontal axis of the angles that determine the positions on the screen where the waves interfere. In the magnetic field such an image would be the variation of the magnetic induction as a function of the positions of the rings centers, where the interaction of the magnetic fields takes place.

- The magnetic field distribution is shown in the histograms of figure (4b) which are similar to the intensity distribution figures in the case of waves [4].

- The diagrams of fringes in both cases present common characteristics such as an enlarged central area and narrower areas around each other. The alternation of the fringes can be due to reinforcement - cancellation or vice versa.

- In wave diffraction, the superposition of the waves from the Huygens points are made on an isophasic wavefront. In our model we consider, the superpositions of the magnetic fields of the rings at the center of the rings. Thus the distance between two successive isophasic wavefronts (the wavelength) which is the wavelength corresponds to the distance d between the centers of the rings. Therefore, the ratio c in both cases expresses the same quantity, so the case $c << 1$ covers both cases for the diffraction to occur.

- At wave diffraction, the randomness that determines the reinforcement and cancellation of the superposition of Huygens waves is due to the different phases corresponding to different paths. In the magnetic field, randomness is introduced through the random numbers that determine the current direction of the rings.

Let us come to our conclusions. Considering a device consisted of a set of similar rings, where their centers lie on an vertical axis to their plane and the currents flowing through the rings have the same value, we detect a continuous transition as the ratio c decreases from the state of a solenoid, where its magnetic induction is given by law of Ampere, to a state where the law is no longer valid and the magnetic induction is independent of the linear density of the rings. A critical value of the ratio c signifying this transition exists (c=1). If we supply this device with randomly direction currents +I, -I, then when we pass the critical value c=1, a pattern of diffraction phenomenon apppears, as in the case of wave propagation in a very narrow obstacle. As c continues to decrease, the diffraction phenomenon of the magnetic field becomes clearer not only as a pattern, but also as a distribution. The diffraction phenomenon monitors the dependence at the magnetic field on the linear density for current of a single direction. Therefore, it is not only due to the randomness of the currents direction, but it is also the result of the theory described above.

A number of similarities between the wave diffraction and the magnetic field behavior in our device reinforce the view that it is in fact the phenomenon of diffusion beyond the wave effects. The investigation towards a universal mechanism to cover this expansion and in other phenomena appears to be an open challenge.

**References**

1. "Magnetic field fluctuations in an array of randomly directed circular currents". V. Anagiannis, Y. Contoyiannis, F. Diakonos. The European Physical Journal B. November 2013, 86; 460.
2. Study on spherical aberration. Correction of solenoid lens in Ultra fast Electron Diffraction. T. Yang, J. J. Li, Y. F. Song, K. J. Fan. 10th Int. Particle Accelerator Conf IPAC2019, Melbourne, Australia JACoW Publishing. ISBN: 978-3-95450-208-0.
3. Kumar, Vinit. "Understanding the focusing of charged particle beams in a solenoid magnetic field." American Journal of Physics 77.8(2009):737.
4. J.D. Jackson, Classical Electrodynamics (Wiley, New York, 1999).
5. David Halliday, Robert Resnick, Kenneth S. Krane, Physics, Volume 2, 5th Edition, ISBN: 978-0-471-40194-0, April 2010.
6. M.Abramowitz, I.A.Stegun, Handbook of Mathematical Functions (Dover, Washington, 1970).

.